\documentclass{article}
\usepackage[utf8]{inputenc}
\usepackage{dsfont}
\usepackage{amssymb}
\usepackage{amsmath}
\usepackage{amsthm}
\usepackage{amsfonts}
\usepackage{mathabx}
\usepackage{enumitem}
\usepackage{hyperref}
\usepackage{graphicx}
\usepackage{tabularx}
\usepackage{caption}
\usepackage{subcaption}
\usepackage[all]{xy}
\usepackage{etoolbox}

\usepackage{qcircuit}

\usepackage{hyperref}

\hypersetup{
    colorlinks=true,
    linkcolor=blue
    }

\title{Quantum Machine Learning for Malware Classification}
\author{Grégoire Barrué and Tony Quertier}
\date{}

\begin{document}
\maketitle

\begin{abstract}
    In a context of malicious software detection, machine    learning (ML) is widely used to generalize to new malware. However, it has been demonstrated that ML models can be fooled or may have generalization problems on malware that has never been seen. We investigate the possible benefits of quantum algorithms for classification tasks. We implement two models of Quantum Machine Learning algorithms, and we compare them to classical models for the classification of a dataset composed of  malicious and benign executable files. We try to optimize our algorithms based on methods found in the literature, and analyze our results in an exploratory way, to identify the most interesting directions to explore for the future.  
\end{abstract}

\section*{Introduction}

Malicious software detection has become an important topic in business, and an important area of research due to the ever-increasing number of successful attacks using malware. In the context of cybersecurity and protection against malware, we face an increase of the different methods that attackers are trying. The attacks are more and more sophisticated and can even be assisted by artificial intelligence \cite{anderson2018learning}, \cite{quertier2022merlin} or by adapted software like Cobalt Strike.

It is thus important to develop techniques that allow to efficiently detect and identify such a massive amount of threats. The use of Machine Learning (ML) can be a good improvement in our methods, and has already proved to be useful \cite{https://doi.org/10.48550/arxiv.2107.11100}\cite{https://doi.org/10.48550/arxiv.2207.02108}\cite{quertier2022merlin}. In our work, we want to investigate the possibilities offered by Quantum Machine Learning (QML), which becomes a research field more and more important for the past few years. It has multiple applications, like for example in quantum chemistry \cite{Reiher_2017}, or again in the estimation of Partial Differential Equation solutions \cite{Kyriienko_2021}. This domain takes origin in quantum computing, and tries to build similar behaviours as classical ML. In our problem, we want to identify whether a certain executable file is a malware or a benign. It is a classification task, and such problems have been studied in a QML framework, for different datasets (MNIST, Iris, ...), as in  \cite{cappelletti2020polyadic},\cite{Chalumuri_2021}, \cite{https://doi.org/10.48550/arxiv.1802.06002}, and with different quantum models. For instance we can refer to Quantum Support Vector Machine (QSVM) \cite{Havl_ek_2019}, Quantum Neural Networks (QNN) \cite{Wan_2017} , or again Quantum Convolutional Neural Networks (QCNN) \cite{Hur_2022}.

More theoretical works begin to arise in QML, in order to optimize the quantum algorithms.
In \cite{Cerezo_2022}, the authors present what has been done in this domain, and what should be prioritized by the near-term research in their opinion. One can also cite \cite{Hoffmann2022}\cite{moussa2022resource}, where some techniques have been developed in order to set up efficient and shot frugal backpropagation, providing alternatives to the use of the parameter shift rule \cite{Schuld_2019}.
Tools to measure the expressivity and the capacity of quantum models are an active research subject \cite{https://doi.org/10.48550/arxiv.2112.04807}\cite{Caro_2022}\cite{Sim_2019}, as the study of symmetries in the data to construct suitable quantum circuits \cite{Meyer2022},\cite{Ragone_2022}.

Our work here is exploratory, and we use some methods found in the literature in order to improve our quantum models for our datasets, while comparing them to classical models. In the next section, we present our datasets, and detail the preprocessings that we use in order to prepare data. Section \ref{sec:QSVM} gathers results for the first type of quantum model that we set up, namely Quantum Support Vector Machine. Then we summarize our results about a second model, a Quantum Neural Network, in Section \ref{sec:QNN}, and present the optimization of this model in Section \ref{sec:optimization}. Finally, we state our conclusions and the future directions for our work. An Appendix is available at the end of the article, gathering the results of all the tests run on our quantum algorithms. 
\section{The two studied datasets}
\label{sec:dataset}
\subsection{Datasets}

For our experiments, we rely on two different datasets,  Bodmas \cite{Yang} and PEMachineLearning \cite{web1}. They all contain malicious and benign Portable Executable (PE) files, distribution and format are summarized in Table \ref{tab:dataset}.

\begin{table}[!ht]
    \centering
     \caption{Distribution and format of each dataset}
    \label{tab:dataset}
    \begin{tabular}{|c|c|c|}
        \hline
         Dataset & Malicious files & Benign files  \\

        \hline
        Bodmas &57,293 & 77,142\\

        \hline
        PEMachine Learning & 114,737 & 86,812 \\
        \hline
    \end{tabular}
\end{table}

Bodmas \cite{Yang} shared with our team a dataset that contains $134,435$ binary files with pre-extracted features together with the $57,293$ malicious files in raw PE format. These files have been collected during one year between August 2019 and September 2020 and labeled: authors indicate the category each file belongs to. The dataset PEMachineLearning contains $201,549$ binary files including $114,737$ malicious files. These files have been gathered from different sources such as VirusShare\footnote{\url{https://virusshare.com/}}, MalShare\footnote{\url{https://malshare.com/}} and TheZoo\footnote{\url{https://github.com/ytisf/theZoo}}. 

\subsection{Preprocessing}

\begin{figure}[!ht]
    \centering
    \begin{subfigure}[!ht]{0.3\textwidth}
        \includegraphics[width=\textwidth]{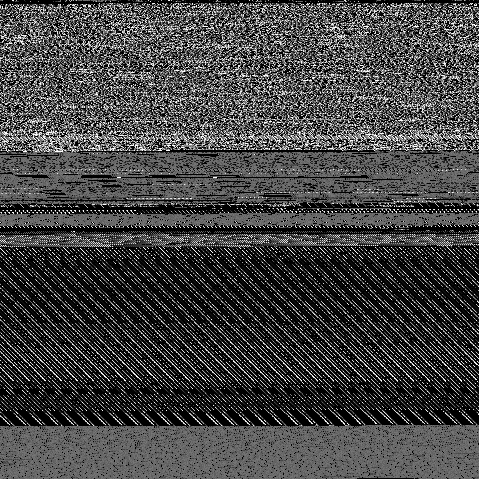}
        \caption{Benign file}
        \label{fig:grayscale_1}
    \end{subfigure}
    \begin{subfigure}[!ht]{0.3\textwidth}
        \includegraphics[width=\textwidth]{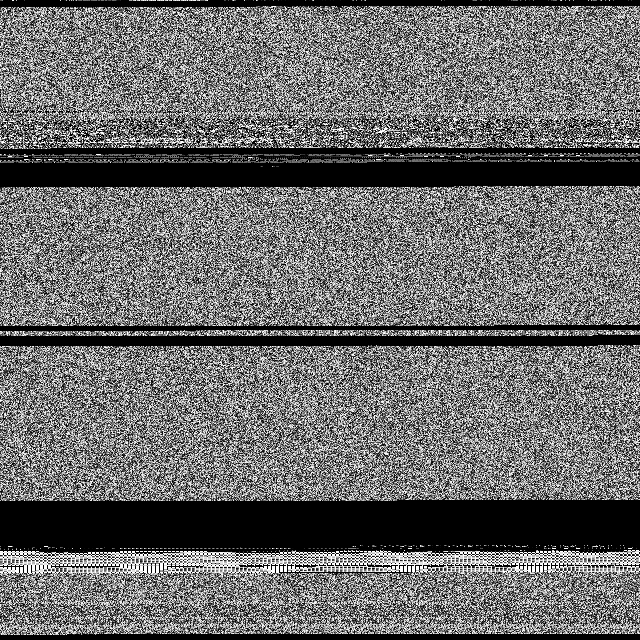}
        \caption{Malicious file}
        \label{fig:grayscale_2}
    
    \end{subfigure}
    \caption{Example of PE files transformation into grayscale images}
    \label{fig:grayscale}
\end{figure}

Training models for malware detection is a multi-step process. The first step consists in extracting information (i.e. features) from the PE files. For this, we rely on two preprocessing algorithms. The first one, a.k.a the Ember method, is detailed in Anderson and al. \cite{Anderson2018} and converts a PE file into a vector of $2,381$ features. Some of these features are listed in Table \ref{tab:Ember}. We need to reduce the number of features for use on a quantum computer. We therefore extract 10 features that we consider as the most impactful and not correlated with each other. For this, we use forward selection, an iterative method which consists in adding the features which improve the most our model. Moreover, we add a feature that we name the \textbf{import maliciousness score} that is calculated according to the confidence level of the imports present in the binary. More information on this topic will soon be available in the PhD thesis of Benjamin Marais. In the following, we denote the preprocessed data in this way as the "Ordered Importance Features" (OIF) dataset.
The second one, a.k.a Grayscale method, was initially submitted by Nataraj and al. \cite{Nataraj2011} and converts a binary file into an image, as can be seen in Figure \ref{fig:grayscale}. To train our models, we choose to resize grayscale images $64\times64$ pixels. In the following, we will denote the preprocessed data in this way as the "Images " dataset.

\begin{table}[!ht]
    \begin{center}
    \caption{Some features computed using Ember extractor}
    \label{tab:Ember}
    \begin{tabular}{|c|c|}
        \hline
         Feature names & Index\\
         \hline
         Byte histogram & 1-256 \\
         Byte entropy & 257-512 \\
         Strings  & 513-616 \\
         General information & 617-626 \\
         Header information & 627-688 \\
         Section & 689-943 \\
         Imports & 944-2223 \\
         Exports & 2224-2350 \\
         Data directory information & 2252 - 2381 \\
         \hline
    \end{tabular}
    \end{center}
\end{table}

\section{Quantum Support Vector Machine}
\label{sec:QSVM}

In this section, we apply a first quantum method known as Quantum Support Vector Machine (QSVM). Our tests are inspired from the Qiskit notebook of the Global Summer School 2021 untitled "Lab 3: Quantum Kernels and Support Vector Machines". We first study the "Images" data set.


First we implement the QSVM with the "ZZFeatureMap", which means that in order to encode our data in the quantum circuits, we use the unitary 
\begin{equation}\label{eq:ZZfeatureMap}    \mathcal{U}_{\phi(x)}=\left(\exp\left(i\sum_{j,k}\phi_{j,k}(x)Z_j\otimes Z_k\right)\exp\left(i\sum_{j}\phi_j(x)Z_j\right)H^{\otimes n}\right)^d.
\end{equation}
In this last equation the components $\phi_{j,k}(x)$ and $\phi_j(x)$ are functions encoding the data, $H$ is the Hadamard gate, $Z$ is the Pauli matrix, and $d$ is the depth of the encoding. Here we choose the number of layers of the feature map $d=2$. The data mapping function $\phi$ is defined by 
\begin{equation}\label{eq:phi_par_defaut}
    \phi_S:x\mapsto \left\{\begin{array}{ll}
       x_j  & \text{ if }S=\{j\} \\
        (\pi-x_j)(\pi-x_k) & \text{ if }S=\{j,k\}
    \end{array}\right..
\end{equation}

\begin{figure}[!h]
    \centering
    \includegraphics[scale=0.6]{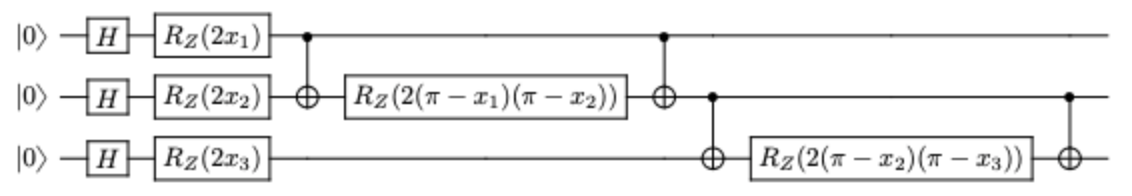}
    \caption{Example of the ZZFeaturemap on three qubits, with $\phi$ defined in \eqref{eq:phi_par_defaut}, linear entanglement and one layer of encoding.}
    \label{fig:ZZFeaturemap}
\end{figure}

Figure \ref{fig:ZZFeaturemap} gives an example of one layer of the ZZfeaturemap, i.e the feature map with $d=1$. The implemented function  is the function $\phi$ defined in \eqref{eq:phi_par_defaut}.
We study the dual form of the classification problem, which means that the algorithm relies on a good estimate of the kernel $K:(x_j,x_k)\mapsto\langle \phi(x_j)\mid\phi(x_k)\rangle$. In order to estimate this kernel, we use quantum implementation: we create a quantum circuit for each entry of the kernel.
We implement our algorithm, and test it for different sizes of data, all from the dataset "Images", and we compare the classification score with other classical algorithms using different kernel estimates, namely linear, polynomial, rbf and sigmoid kernels. We also compute the F1-score of our algorithm, which is defined as the harmonic mean of precision and recall. We gather the results in  Table \ref{fig:tab_acc_QSVM/SVMs_data_img}.

\begin{table}[!h]
    \centering
    \begin{tabular}{|c|c|c|c|c|c|c|}
  \hline
    data (train/test) & QSVM & linear & poly & rbf & sigmoid & F1-score \\
  \hline
    500/100 & 0.78 & 0.77 & 0.78 & 0.80 & 0.61 & 0.778 \\
    1000/200 & 0.815 & 0.8 & 0.81 & 0.79 & 0.55 & 0.838 \\
    2000/400 & 0.8125 & 0.77 & 0.78 & 0.8 & 0.54 & 0.827 \\
    4000/800 & 0.7925 & 0.78 & 0.80 & 0.82 & 0.55 & 0.805 \\
    8000/1600 & 0.805 & 0.77 & 0.79 & 0.82 & 0.55 & 0.811 \\
    16000/4000 & 0.81075 & 0.77 & 0.78 & 0.81 & 0.56 & 0.814 \\
  \hline
\end{tabular} \\
    \caption{Accuracy comparison between the QSVM and several classical SVMs, for different numbers of train and test data samples. The studied dataset is "Images"}
    \label{fig:tab_acc_QSVM/SVMs_data_img}
\end{table}

Here we already see that on our dataset, we have comparable, and even sometimes slightly better results with the quantum model than with the classical SVMs. 
We can also customise the Pauli gates in the feature map, for example, $P_0 = X, P_1 = Y, P_2 = ZZ$:
\begin{equation}
\begin{split}
    \mathcal{U}_{\Phi(\mathbf{x})} = \left( \exp\left(i\sum_{jk} \phi_{\{j,k\}}(\mathbf{x})  Z_j \otimes Z_k\right) \right. &  \exp\left(i\sum_{j} \phi_{\{j\}}(\mathbf{x}) Y_j\right) \\ &\left. \exp\left(i\sum_j \phi_{\{j\}}(\mathbf{x}) X_j\right) H^{\otimes n} \right)^d.
\end{split}    
\end{equation}
$$$$ 

For this feature map, we still take $d=2$, which gives us a much deeper circuit, because we use more elementary gates to implement one layer of this map than for one layer of the ZZFeatureMap. We refer to this map as the PauliFeaturemap.

Another possibility is to change the data mapping function $\phi$, and take for example $$\phi_S:\mathbf{x}\mapsto \Bigg\{\begin{array}{ll}
    x_i & \mbox{if}\ S=\{i\} \\
        \sin(\pi-x_i)\
        sin(\pi-x_j) & \mbox{if}\ S=\{i,j\}
    \end{array}.$$
We encode this data mapping function with the ZZFeaturemap defined in Equation \eqref{eq:ZZfeatureMap}. We refer to this map as the ZZphiFeaturemap.

Finally, we implement a simpler feature map, which is the ZFeatureMap : 
\begin{equation}
    \mathcal{U}_{\phi(x)}=\left(\exp\left(i\sum_j\phi_{j}(x)Z_j\right)H^{\otimes n}\right)^d.
\end{equation}
Here we take $\phi$ as defined in Equation \eqref{eq:phi_par_defaut}. This encoding does not have entanglement, which means that it is classically computable, then no quantum advantage should be observed. However it is interesting to check if the other models, which have entanglement, does perform better than this one in particular. 
The results of all these models are showed in Table \ref{fig:tab_acc_QSVMs_data_img}, in term of accuracy score. 
\begin{table}[!h]
    \centering
   \begin{tabular}{|c|c|c|c|c|}
  \hline
    data (train/test) & ZZFeature & PauliFeature & ZZphiFeature & ZFeature \\
  \hline
    500/100 & 0.78 & 0.75 & 0.82 & 0.81  \\
    1000/200 & 0.815 & 0.79 & 0.815 & 0.805  \\
    2000/400 & 0.8125 & 0.79 & 0.8075 & 0.8  \\
    4000/800 & 0.7925 & 0.786 & 0.811 & 0.8075  \\
    8000/1600 & 0.805 & 0.7975 & 0.814 & 0.808  \\
    16000/4000 & 0.81075 & 0.806 & 0.805 & 0.799 \\
  \hline
\end{tabular}
    \caption{Accuracy comparison between QSVMs with different feature maps, for different numbers of train and test data samples, on the dataset "Images"}
    \label{fig:tab_acc_QSVMs_data_img}
\end{table}

Then, we investigate the influence of the dimension of the PCA in our results. This dimension corresponds to the number of features that we use in our models. In the quantum case, it also corresponds to the number of qubits that are necessary to encode our data. Table \ref{fig:tab_acc_QSVMs/SVMs_PCA_img} gathers the accuracy scores between the different quantum algorithms and the different classical algorithms, for different numbers of features. Finally, we also gather the F1-score for the different quantum algorithms, according to the number of features, in Table \ref{fig:tab_F1_QSVMs_PCA_img}. 

\begin{table}[!h]
    \centering
    \begin{tabular}{|c|c|c|c|c|}
     \hline
     PCA(train/test) & 3(2000/400) & 4(2000/400) & 6(1000/200) & 7(1000/200)  \\
     \hline
     ZZFeature & 0.7675 & 0.8125 & 0.81 & 0.825 \\
     PauliFeature & 0.7575 & 0.79 & 0.715 & 0.765 \\
     ZZphiFeature & 0.8 & 0.8075 & 0.79 & 0.785 \\
     ZFeature & 0.7825 & 0.8 & 0.785 & 0.78 \\
     linear & 0.79 & 0.77 & 0.79 & 0.79 \\
     poly & 0.77 & 0.78 & 0.80 & 0.80 \\
     rbf & 0.81 & 0.8 & 0.8 & 0.8 \\
     sigmoid & 0.42 & 0.54 & 0.53 & 0.57\\
     \hline
\end{tabular}\\
    \caption{Accuracy comparison between different quantum and classical SVMs, for different values of PCA, on the dataset "Images". }
    \label{fig:tab_acc_QSVMs/SVMs_PCA_img}
\end{table}

\begin{table}[!h]
    \centering
    \begin{tabular}{|c|c|c|c|c|}
  \hline
    PCA (train/test) & ZZFeature & PauliFeature & ZZphiFeature & ZFeature \\
  \hline
    3(2000/400) & 0.79 & 0.774 & 0.819 & 0.804  \\
    4(2000/400) & 0.827 & 0.814 & 0.826 & 0.819  \\
    6(1000/200) & 0.828 & 0.739 & 0.817 & 0.815  \\
    7(1000/200) & 0.844 & 0.791 & 0.810 & 0.810  \\
  \hline
\end{tabular}\\
    \caption{F1-score comparison between the different QSVMs, for different values of PCA, on the dataset "Images".}
    \label{fig:tab_F1_QSVMs_PCA_img}
\end{table}

With all these results, we see that for low PCA the quantum algorithms does not provide improvements compared to classical algorithms, for small and big datasets. However, for a rather small dataset, namely 1000 or 2000 samples, we can see that increasing the dimension of the PCA allows to get a better accuracy with the quantum circuit based on the ZZFeatureMap than with any classical algorithm. In particular, if we take a PCA dimension of 7, we observe a gain of $2.5\%$ of accuracy. Our intuition is thus that quantum algorithms could perform better than classical algorithms on small datasets. In some problems, such as many use-cases in cybersecurity, there is limited available data and it is interesting to extract as much information as possible from the data. The main issue would then be, as usual, to adapt the algorithms according to the number of available qubits.

Table \ref{fig:tab_acc_QSVMs/SVMs_IBM_img} gathers the different QSVMs, tested on the real IBM hardware, compared to classical SVMs, for a reduced dataset composed of 100 train samples and 20 test samples. 

\begin{table}[!h]
    \centering
    \begin{tabular}{|c|c|c|}
   \hline
   SVM  & accuracy & F1-score \\
   \hline
   ZZFeature & 0.75 & 0.8 \\
   PauliFeature & 0.8 & 0.83 \\
   ZZphiFeature & 0.55 & 0.709 \\
   ZFeature & 0.75 & 0.78 \\
   linear & 0.85 & / \\
   poly & 0.8 & / \\
   rbf & 0.75 & / \\
   sigmoid & 0.85 & / \\
   \hline 
\end{tabular}\\
    \caption{Accuracy comparison on the real IBM hardware between the QSVMs and the classical SVMs, on the dataset "Images". }
    \label{fig:tab_acc_QSVMs/SVMs_IBM_img}
\end{table}

We can see that testing our models on the IBM device gives us results which are similar to classical results. Note here that for practical reasons, namely because of the computation time and the fact that we have not full access to IBM device yet (which implies queues between each computation), this test is made with a very little training set and test set. Thus the difference between quantum and classical models represents only an error for two or three labels.  

We also run our tests on the OIF dataset. The features are ordered by importance. The first feature is a malicious score, which determines the point at which the imports present in the files are dangerous. Table \ref{fig:tab_acc_QSVMS/SVMs_data_features} gathers the accuracy scores of both quantum and classical SVMS, for different sizes of train and test sets.

\begin{table}[!h]
    \centering
    \includegraphics[scale=0.68]{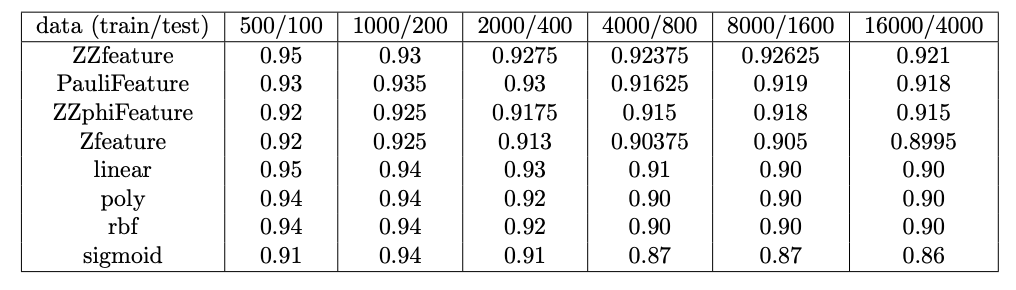}
    \caption{Accuracy comparison between QSVMs and SVMs for different numbers of train and test data samples, on the dataset OIF.}
    \label{fig:tab_acc_QSVMS/SVMs_data_features}
\end{table}

The F1-scores of the different QSVMs are presented in Table \ref{fig:tab_F1_QSVMs_data_feature}.
As for the previous dataset, we want to investigate the influence of the number of features on the performances of the models. The results are shown in Table \ref{fig:tab_acc_QSVMs/SVMs_PCA_features}.
\begin{table}[!h]
    \centering
    \includegraphics[scale=0.6]{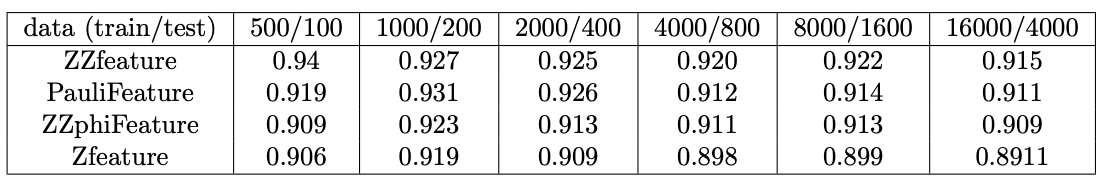}
    \caption{F1-score comparison between different QSVMs for different numbers of train and test data samples, on the dataset OIF. }
    \label{fig:tab_F1_QSVMs_data_feature}
\end{table}

\begin{table}[!h]
    \centering
    \includegraphics[scale=0.75]{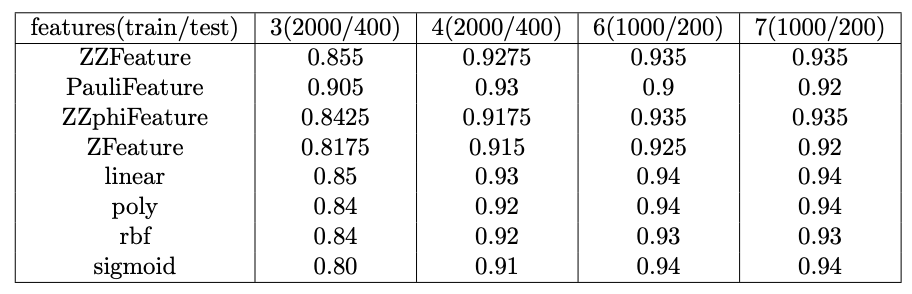}
    \caption{Accuracy comparison between different QSVMs and classical SVMs for different numbers of features, on the dataset OIF.}
    \label{fig:tab_acc_QSVMs/SVMs_PCA_features}
\end{table}

Here, we see that for small numbers of features, some quantum models perform better than classical models. 
On this dataset, the features are ordered by importance, thus Table \ref{fig:tab_acc_QSVMs/SVMs_PCA_features} gives us the intuition that quantum model extract the most of the information from the first features, while classical model seem to gain in accuracy more uniformly when we add features. 
Finally, we also gather the F1-score for the different quantum algorithms, according to the number of features, in Table \ref{fig:tab_F1_QSVMs_PCA_features}. 

\begin{table}[!h]
    \centering
    \begin{tabular}{|c|c|c|c|c|}
  \hline
    qubits(train/test) & ZZFeature & PauliFeature & ZZphiFeature & ZFeature \\
  \hline
    3(2000/400) & 0.859 & 0.925 & 0.933 & 0.931  \\
    4(2000/400) & 0.901 & 0.926 & 0.90 & 0.917  \\
    6(1000/200) & 0.843 & 0.913 & 0.932 & 0.932  \\
    7(1000/200) & 0.814 & 0.909 & 0.919 & 0.914  \\
  \hline
\end{tabular}\\
    \caption{F1-score comparison between different QSVMs for different numbers of qubits, on the dataset OIF.}
    \label{fig:tab_F1_QSVMs_PCA_features}
\end{table}

First of all, we remark that the accuracy scores and F1-scores with this dataset are much higher than with the dataset "Images". It could be explained by the fact that here the features are carefully chosen and contain a lot of information about the data. Thus, even with a few features we capture most of the information, whereas in the first case we use a PCA to reduce the number of features, which does not give us information about the new features that are created. 
On the little datasets we have similar scores or slightly better scores for the classical models, but since there is few samples, the difference between the models represents only a few labels. Besides, when we increase the number of data, the ratio is not the same and this time quantum SVMs perform better than classical ones, with around $1.5\%$ more. 
Furthermore, note that if one increases the number of features/qubits, the accuracy also increases. However, this case is different than the previous one where the quantum advantage appears when we have more qubits. Indeed, here the quantum algorithms seem to perform better than the classical ones when we have less features, namely the most important ones. 
Finally, we would like to add that since there are still several improvements that can be made for the Qiskit simulator (like the fact that the computations of our quantum circuits are not parallelized), the quantum algorithms need a lot more computation time than classical algorithms, which limits their efficiency and our experiment. Testing them on a bigger dataset, or using more qubits, would take too much time without any guarantee of a quantum advantage, thus we prefer to run them on smaller datasets or fewer qubits in order to have more results to analyze. It is relevant since not  many logical qubits are available on real quantum hardware yet.

\section{Quantum Neural Network}
\label{sec:QNN}
\subsection{Presentation of our framework}

In this section, we still investigate the classification problem from our two datasets, but this time we use a different architecture of quantum algorithm. Here we want to use a Quantum Neural Network (QNN), which is a parameterized quantum circuit, with parameters encoding the input data, and others trained by a gradient descent. The principle of this type of quantum models is very similar to classical neural networks \cite{Wan_2017}, hence their denomination.
We implement the quantum algorithm presented in \cite{Chalumuri_2021}, adapted for a two-class problem. This algorithm has the advantage of being well-detailed and adaptable for multiple class problems as well. In addition of having been tested on real quantum hardware provided by IBM, at least for the state preparation. 

\begin{figure}[!h]
    \centering
    \includegraphics[scale=0.57]{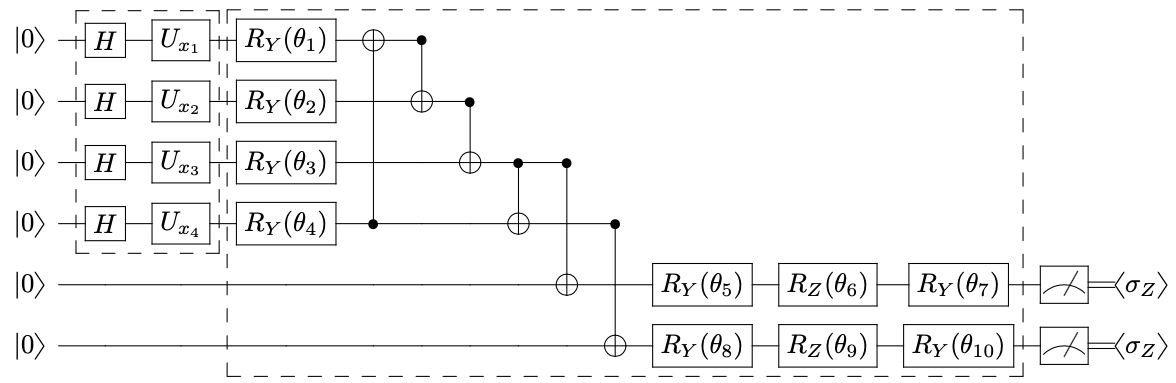}
    \caption{Example of the quantum circuit, for one layer and four qubits. The  first smaller box represents the data encoding, while the second bigger box represents a layer of the circuit. }    
    \label{Circuit quantique}
\end{figure}

More precisely, the state preparation consists in sending all classical input data $x$ in $[0,\frac{\pi}{2}]^N$, where $N$ is the number of  data qubits. Then, classical data is transformed into quantum data $|x\rangle$ where each coordinate of $|x\rangle$ is in the state $\cos(x_j)|0\rangle + \sin(x_j)|1\rangle$. It is done applying on each data qubit a Hadamard gate and then applying on qubit $j$ the gate $U(x_j)$ defined by $$U(x_j)=\begin{pmatrix}
\cos(\frac{\pi}{4}-x_j) & \sin(\frac{\pi}{4}-x_j)\\
-\sin(\frac{\pi}{4}-x_j) & \cos(\frac{\pi}{4}-x_j) \end{pmatrix}.$$
To the $N$ data qubits we add two ancilla qubits, because we limit ourselves to a two-class problem. Once the initial state is prepared, we add layers, which play the same role as hidden layers in a  classical Neural Network algorithm. A layer $\ell$ consists in applying on each data qubit $j$ a gate $R_Y(\theta_j^\ell)$, with $\theta_j^\ell\in\Theta$ the parameter space, then entangling all data qubits and ancilla qubits thanks to CNOT gates. Next, we apply on each ancilla qubit a sequence of three parameterized rotations : according to Y, Z, and again Y. Finally, we take the expectation of each ancilla qubit in the Z-Pauli basis. 
The parameter optimization is made by minimizing the cross entropy of the system, using the parameter shift rule. For more details one can read \cite{Chalumuri_2021}. 
Figure \ref{Circuit quantique} illustrates the quantum circuit that we use, in the case of 4 qubits and one layer of parameters.

\subsection{Numerical experiments}

In order to lighten the lecture of this section, we choose to let the most part of the tables gathering our results in the Appendix, and refer to them while analyzing the different models or datasets that we study.

First we choose to train a naive version of our model, namely we use the model as described, and we set up the parameter shift rule for the gradient descent (\cite{Schuld_2019}). The advantage of this method is that it is implementable directly on quantum hardware, at the cost of using the quantum circuit three times per  trainable parameter, which could be a problem for deep circuits.
Once the learning of the quantum algorithms done, we measure their accuracy. According to the intuition given by the results of the QSVM in the previous section, and because processing lots of data would make the computations a lot much longer due to parameter shift rule, we fix the dataset to only 1000 training samples and 200 testing samples. We still can play with some parameters, as the learning rate, the PCA dimension (thus the number of data qubits), the number of layers, or again the number of epochs. We do not make different batch sizes, because according to \cite{Hoffmann2022} it should not impact the performances of an algorithm relying on parameter shift rule. First we fix the learning rate to $0.1$, make only one epoch, and investigate the influence of the number of layers, for the algorithm with 4 and 7 data qubits. Thus we make the number of layers vary from 1 to 3, and compute the accuracy and the learning time for every case. We obtain  Table \ref{fig:tab_acc_QNN_layers_img200}. 

\begin{table}[!h]
    \centering
    \begin{tabular}{|c|c|c|c|c|c|c|}
  \hline
    qubits/layers & 4/1 & 4/2 & 4/3  \\
  \hline
    accuracy (train/test) & 0.584/0.74 & 0.609/0.605 & 0.606/0.74    \\
    learning time & 9min44s & 26min24s & 51min24s  \\
  \hline
  qubits/layers & 7/1 & 7/2 & 7/3 \\
  \hline
  accuracy (train/test) & 0.537/0.51 & 0.513/0.445 & 0.539/0.545 \\
  learning time & 17min7s & 42min33s & 1h23min \\
  \hline
\end{tabular}\\
    \caption{Accuracy scores of the QNN, for different numbers of qubits, and different numbers of layers, on the dataset "Images" for 200 test samples.}
    \label{fig:tab_acc_QNN_layers_img200}
\end{table}

\begin{table}[!h]
    \centering
    \begin{tabular}{|c|c|c|c|c|c|c|}
  \hline
    PCA/layers & 4/1 & 4/2 & 4/3 \\
  \hline
    accuracy (train/test) & 0.489/0.33 & 0.489/0.410 & 0.492/0.4   \\
  \hline
  PCA/layers &  7/1 & 7/2 & 7/3 \\
  \hline
  accuracy (train/test) & 0.489/0.420 & 0.489/0.435 & 0.492/0.475 \\
  \hline
\end{tabular}\\
    \caption{Accuracy scores of the classical NN, for different PCA values and different numbers of layers, on the dataset "Binaire.images" for 200 test samples.}
    \label{fig:tab_acc_NN_layers_img200}
\end{table}

We also present in Table \ref{fig:tab_acc_NN_layers_img200} a similar table but for a classical neural network, the number of qubits being equivalent to the dimension of the PCA.
We then make other tests. First we fix the number of layers to 2, with only one epoch, and investigate the influence of the learning rate $\eta$ by running the algorithm for three different values, namely $\eta=0.01$, $\eta=0.1$ and $\eta=0.5$. For each learning rate value, we run the model on 4 and 7 qubits, and we compare the results with the same test on a classical NN. The results are presented in the Appendix, Tables \ref{fig:tab_acc_QNN_learning_img200} and \ref{fig:tab_acc_NN_learning_img200}. 






Once these tests on shallow circuits done, we try to observe what could happen on deeper circuits, to see for example if we can observe similar behaviours as for classical neural networks, and to try to extract an optimal model for our problem. Thus we try three different circuit and their classical equivalent. First we implement a quantum circuit with 4 qubits, 5 layers and 10 epochs, with a learning rate of 0.05. Then we increase the number of layers to 10, to see if a deeper circuit performs better, this gives us Tables \ref{fig:tab_acc_QNN_4/10/10_img200}   and \ref{fig:tab_acc_NN_4/10/10_img200}  in Appendix. Finally, we try to look for an influence of the number of qubits, and run a quantum circuit with 7 qubits, 5 layers and 5 epochs, with a learning rate of $0.1$. The results of this last model are presented in the Appendix in Tables \ref{fig:tab_acc_QNN_7/5/5_img200} and \ref{fig:tab_acc_NN_7/5/5_img200}.
In this section we present the first algorithm, with 4 qubits, 5 layers and 10 epochs. The accuracy scores, epoch by epoch are presented in Table \ref{fig:tab_acc_QNN_4/5/10_img200} for the quantum model and Table \ref{fig:tab_acc_NN_4/5/10_img200} for its classical equivalent.



\begin{table}[!h]
    \centering
    \begin{tabular}{|c|c|c|c|c|c|c|}
  \hline
     & epoch 1 & epoch 2 & epoch 3 & epoch 4 & epoch 5 \\
  \hline
    4 qubits/5 layers & 0.58 & 0.59 & 0.59 & 0.62 & 0.63   \\
  \hline
\end{tabular}\\

\begin{tabular}{|c|c|c|c|c|c|c|}
  \hline
     & epoch 6 & epoch 7 & epoch 8 & epoch 9 & epoch 10 & test \\
  \hline
    4 qubits/5 layers & 0.63 & 0.62 & 0.64 & 0.63 & 0.62 & 0.64  \\
  \hline
\end{tabular}\\
    \caption{Accuracy scores of the QNN, for each epoch, with 4 qubits and 5 layers, on the dataset "Images" for 200 test samples.}
    \label{fig:tab_acc_QNN_4/5/10_img200}
\end{table}


\begin{table}[!h]
    \centering
    \begin{tabular}{|c|c|c|c|c|c|c|}
  \hline
     & epoch 1 & epoch 2 & epoch 3 & epoch 4 & epoch 5 \\
  \hline
    PCA 4/5layers & 0.5040 & 0.5150 & 0.477 & 0.500 & 0.5250   \\
  \hline
\end{tabular}\\

\begin{tabular}{|c|c|c|c|c|c|c|}
  \hline
     & epoch 6 & epoch 7 & epoch 8 & epoch 9 & epoch 10 & test \\
  \hline
    PCA 4/5layers & 0.5260 & 0.5020 & 0.5040 & 0.4910 & 0.4870 & 0.5400  \\
  \hline
\end{tabular}\\
    \caption{Accuracy scores of the classical NN, for each epoch, with PCA 4 and 5 layers, on the dataset "Images" for 200 test samples.}
    \label{fig:tab_acc_NN_4/5/10_img200}
\end{table}






First, let us note that the results are much less satisfying than with the QSVMs. Here we seem to have a better accuracy on the QNN than on the classical neural network, but this classical algorithm is run with a batch-size of 1 in order to create exactly the same framework as for the QNN. We know that it is not realistic to take such a batch-size for a classical neural network, but we wanted to have a naive comparison at first.
We seem to perform better with 4 qubits than with 7. If we analyze the influence of the layers, we can see that the accuracy increases with this number, as in the classical case. However, the increasing is not very significant, which could raise some problems because lots of parameters would be necessary in order to get a decent accuracy for this dataset. Finally, adding several epochs also increases the accuracy, but we can see that we do not need lots of epochs, because the accuracy rapidly stagnates, namely after two or three epochs. This is good news, because it limits the necessary depth of the model: we do not need as much epochs in order to reach a certain accuracy score, or at least it is not the solution to increase this number. 
The influence of the learning rate in this case does not give any conclusion, we could investigate in this direction, running tests for more learning rates. 
Finally, the accuracy scores measured for the learning phases and for the testing phases are coherent, we do not observe overfitting phenomenon. 

Next we focus on the same dataset, but this time we train the model with 1000 samples, and we test it on 4000 samples. We omit the learning time, and we run the same tests. 
In  Table \ref{fig:tab_acc_QNN_layers_img}, we use only 1 epoch and a learning rate $\eta=0.1$. We also present its equivalent on a classical neural network in Table \ref{fig:tab_acc_NN_layers_img}.
The influence of the learning rate (Tables \ref{fig:tab_acc_QNN_learning_img} and \ref{fig:tab_acc_NN_learning_img}), the accuracy scores of the model with 4 qubits, 10 layers, 10 epochs, $\eta=0.05$ (Tables \ref{fig:tab_acc_QNN_4/10/10_img} and \ref{fig:tab_acc_NN_4/10/10_img}) and those of the circuit with 7 qubits, 5 layers, 5 epochs and $\eta=0.1$ (Tables \ref{fig:tab_acc_QNN_7/5/5_img} and \ref{fig:tab_acc_NN_7/5/5_img}) are presented in the Appendix. 
We present on Table \ref{fig:tab_acc_QNN_4/5/10_img} the accuracy scores for the quantum circuit composed of 4 qubits, 5 layers, 10 epochs and a learning rate of $0.05$. The accuracy scores of the same test on a classical NN are available in Table \ref{fig:tab_acc_NN_4/5/10_img}.

\begin{table}[!h]
    \centering
    \begin{tabular}{|c|c|c|c|c|c|c|}
  \hline
    qubits/layers & 4/1 & 4/2 & 4/3  \\
  \hline
    accuracy (train/test) & 0.5/0.518 & 0.515/0.508 & 0.542/0.512   \\
  \hline
  qubits/layers & 7/1 & 7/2 & 7/3 \\
  \hline
  accuracy (train/test) & 0.513/0.497 & 0.537/0.5077 & 0.526/0.5125 \\
  \hline 
\end{tabular}\\
    \caption{Accuracy scores of the QNN, for different numbers of qubits and different numbers of layers, on the dataset "Images". }
    \label{fig:tab_acc_QNN_layers_img}
\end{table}

\begin{table}[!h]
    \centering
    \begin{tabular}{|c|c|c|c|c|c|c|}
  \hline
    PCA/layers & 4/1 & 4/2 & 4/3  \\
  \hline
    accuracy (train/test) & 0.6570/0.761 & 0.668/0.630 & 0.677/0.6575   \\
  \hline
  PCA/layers & 7/1 & 7/2 & 7/3 \\
  \hline
  accuracy (train/test) & 0.684/0.779 & 0.663/0.5573 & 0.703/0.7638 \\
  \hline
\end{tabular}\\
    \caption{Accuracy scores of the classical NN, for different PCA values and different numbers of layers, on the dataset "Images".}
    \label{fig:tab_acc_NN_layers_img}
\end{table}

\begin{table}[!h]
    \centering
    \begin{tabular}{|c|c|c|c|c|c|c|}
  \hline
     & epoch 1 & epoch 2 & epoch 3 & epoch 4 & epoch 5 \\
  \hline
    4 qubits/5layers & 0.50 & 0.55 & 0.53 & 0.52 & 0.53   \\
  \hline
\end{tabular}\\

\begin{tabular}{|c|c|c|c|c|c|c|}
  \hline
     & epoch 6 & epoch 7 & epoch 8 & epoch 9 & epoch 10 & test \\
  \hline
    4 qubits/5layers & 0.56 & 0.57 & 0.56 & 0.55 & 0.54 & 0.55  \\
  \hline
\end{tabular}\\
    \caption{Accuracy scores of the QNN, for each epoch, with 4 quibts and 5 layers, on the dataset "Images".}
    \label{fig:tab_acc_QNN_4/5/10_img}
\end{table}


\begin{table}[!h]
    \centering
    \begin{tabular}{|c|c|c|c|c|c|c|}
  \hline
     & epoch 1 & epoch 2 & epoch 3 & epoch 4 & epoch 5 \\
  \hline
    PCA 4/5layers & 0.7090 & 0.7150 & 0.7500 & 0.7380 & 0.7290   \\
  \hline
\end{tabular}\\

\begin{tabular}{|c|c|c|c|c|c|c|}
  \hline
     & epoch 6 & epoch 7 & epoch 8 & epoch 9 & epoch 10 & test \\
  \hline
    PCA 4/5layers & 0.7240 & 0.7580 & 0.7490 & 0.7550 & 0.7580 & 0.7333  \\
  \hline
\end{tabular}\\
    \caption{Accuracy scores of the classical NN, for each epoch, with PCA 4 and 5 layers, on the dataset "Images".}
    \label{fig:tab_acc_NN_4/5/10_img}
\end{table}

In this second series of tests, we have similar accuracy scores, with more data to classify, using the same learning size. Here the influence of the learning rate for 4 qubits seems to be like in \cite{Hoffmann2022}, but not for 7 qubits, which is not very conclusive.
Here we see much better results on the classical neural network, where this time we take a batch-size of 16 to be more realistic, than on the quantum algorithm. However, in both classical and quantum cases, we do not observe any advantage in taking 7 qubits rather than 4 qubits.
We also can see in a much clearer way the fact that increasing the number of epochs is not very useful: in the case of 4 qubits, 10 layers and 10 epochs, the algorithm does not learn more after two epochs. It is slightly better with the two other tests (4/5/10 and 7/5/5) but we do not observe a relevant increasing through the epochs.
We observe very few differences between 5 and 10 layers. One explication might be that in this case the overparametrization regime is reached for (or before) 5 layers, so that it is not necessary to add more layers (see \cite{Larocca_2021}). As a reminder, overparametrization is the fact that beyond a critical number of parameters, the optimizer converges exponentially fast to solution. It could be interesting to verify this hypothesis with more theoretical computations, namely computing the dynamical Lie algebra of the model and its Quantum Fisher Information Matrices. 

We run the same test on a different dataset, where we work directly on the features of the files. We class these features by importance and relevance, and we also add a malicious rate as a additional feature. Here are the results, for the same parameters as before, presented in Table \ref{fig:tab_acc_QNN_layers_features} for the quantum model and in Table \ref{fig:tab_acc_NN_layers_features} for the classical neural network.

\begin{table}[!h]
    \centering
    \begin{tabular}{|c|c|c|c|c|c|c|}
  \hline
    qubits/layers & 4/1 & 4/2 & 4/3  \\
  \hline
    accuracy (train/test) & 0.698/0.667 & 0.677/0.667 & 0.803/0.819   \\
  \hline
  qubits/layers & 7/1 & 7/2 & 7/3 \\
  \hline
  accuracy (train/test) & 0.697/0.784 & 0.487/0.504 & 0.802/0.804 \\
  \hline
\end{tabular}\\
    \caption{Accuracy scores of the QNN for different numbers of qubits and different numbers of layers, on the dataset OIF.}
    \label{fig:tab_acc_QNN_layers_features}
\end{table}


\begin{table}[!h]
    \centering
        \begin{tabular}{|c|c|c|c|c|c|c|}
  \hline
    features/layers & 4/1 & 4/2 & 4/3 \\
  \hline
    accuracy (train/test) & 0.4890/0.1965 & 0.8370/0.5910 & 0.848/0.7588   \\
  \hline
  features/layers & 7/1 & 7/2 & 7/3 \\
  \hline
  accuracy (train/test) & 0.8440/0.5545 & 0.829/0.5312 & 0.8410/0.7377 \\
  \hline
\end{tabular}
    \caption{Accuracy scores of the classical NN, for different numbers of qubits and different numbers of layers, on the dataset OIF.}
    \label{fig:tab_acc_NN_layers_features}
\end{table}

The influence of the learning rate (Tables \ref{fig:tab_acc_QNN_learning_features} and \ref{fig:tab_acc_NN_learning_features}), the accuracy scores of the model with 4 qubits, 10 layers, 10 epochs, $\eta=0.05$ (Tables \ref{fig:tab_acc_QNN_4/10/10_features} and \ref{fig:tab_acc_NN_4/10/10_features}) and those of the circuit with 7 qubits, 5 layers, 5 epochs and $\eta=0.1$ (Tables \ref{fig:tab_acc_QNN_7/5/5_features} and \ref{fig:tab_acc_NN_7/5/5_features}) are presented in the Appendix.
We present in Table \ref{fig:tab_acc_QNN_4/5/10_features} the accuracy scores of our quantum model with 4 qubits, 5 layers and 10 epochs, for a learning rate $\eta=0.05$. Furhtermore, Table \ref{fig:tab_acc_NN_4/5/10_features} presents the scores of the equivalent classical NN.


\begin{table}[!h]
    \centering
    \begin{tabular}{|c|c|c|c|c|c|c|}
  \hline
     & epoch 1 & epoch 2 & epoch 3 & epoch 4 & epoch 5 \\
  \hline
    4 qubits/5layers & 0.796 & 0.808 & 0.809 & 0.809 & 0.809   \\
  \hline
\end{tabular}\\

\begin{tabular}{|c|c|c|c|c|c|c|}
  \hline
     & epoch 6 & epoch 7 & epoch 8 & epoch 9 & epoch 10 & test \\
  \hline
    4 qubits/5layers & 0.809 & 0.809 & 0.809 & 0.809 & 0.809 & 0.81  \\
  \hline
\end{tabular}
    \caption{Accuracy scores of the QNN, for each epoch, with 4 qubits and 5 layers, on the dataset OIF.}
    \label{fig:tab_acc_QNN_4/5/10_features}
\end{table}


\begin{table}[!h]
    \centering
    \begin{tabular}{|c|c|c|c|c|c|c|}
  \hline
     & epoch 1 & epoch 2 & epoch 3 & epoch 4 & epoch 5 \\
  \hline
    4 features/5layers & 0.86 & 0.863 & 0.883 & 0.881 & 0.885   \\
  \hline
\end{tabular}\\

\begin{tabular}{|c|c|c|c|c|c|c|}
  \hline
     & epoch 6 & epoch 7 & epoch 8 & epoch 9 & epoch 10 & test \\
  \hline
    4 features/5layers & 0.884 & 0.872 & 0.887 & 0.89 & 0.894 & 0.847  \\
  \hline
\end{tabular}\\
    \caption{Accuracy scores of the classical NN, for each epoch, with 4 features and 5 layers, on the dataset OIF.}
    \label{fig:tab_acc_NN_4/5/10_features}
\end{table}


As for the QSVM case, we have much better results on this dataset than on the dataset "Images". This could also be due to the fact that images of benigns and malware are white noise images, so that they may be hard to identify, especially with the PCA method in addition, which mixes the features of these images.
Here again, we do not see a big difference between 4 or 7 qubits, which here could be explained by the fact that the first features, which are considered in both cases, represent most of the information of the data. Thus the quantum model seems to extract lots of information from the data, even with a small number of features. However we see an increasing of the accuracy with the number of layers, at least up to three layers. It was already the case on the previous dataset but here it is more obvious.  For more layers it becomes more subtil, raising again the question of overparametrization: maybe the model cannot reach higher rates of accuracy, thus it is useless to add layers of trainable parameters.
We also see the increasing of the accuracy with the increasing of the learning rate, which correspond to the remarks in \cite{Hoffmann2022}. We know that we have just tested with three values, and that we should run tests with more values of learning rates to give a real conclusion.
Even in this case where we have good accuracy scores, the classical neural network with a batch-size of 16 performs better, but here the difference between the two algorithms is smaller, especially for deeper models.

For a quantum neural network, or at least for our model, we note that it does not seem relevant to play with the number of qubits (corresponding to the number of features), but it is very likely to be because most information is contained in the first features for our dataset. We do not observe a difference in the classical case neither. Here again, some of the computations could be parallelized, because the parameter shift rule makes us use the quantum circuit three times by parameter. Thus every test that we run takes us more time, so that we chose not to run too deep circuits, in order to obtain more results. 

All these results are interesting, but yet not very convincing when compared for instance to the QSVM model. We could conclude that in order to solve our problem, the QNN architecture (or at least our architecture) is not really adapted. To get a clearer opinion on this question, we want to optimize our algorithm, with a couple of methods that are detailed on the next section. 




\section{Optimization of the QNN}
\label{sec:optimization}
\subsection{SPSB}
In this section we try to optimize the QNN using a method developed in \cite{Hoffmann2022}, called SPSB (Simultaneous Perturbation for Stochastic Backpropagation). This method allows to dramatically reduce the number of quantum circuit evaluations, estimating quantum gradients in random directions and using Automatic differentiation (or backpropagation). Thus batches are necessary in order to average the random directions and obtain a coherent gradient descent. In this method, we do not have to compute each coordinate of the gradient, which in the parameter shift rule demands to use three quantum circuit evaluations. Here the learning times does not depend on the number of parameters anymore, because we pass in the quantum circuit three times by input data, regardless of the depth of the circuit. This method is based on the finite difference scheme $$\partial_x f(x)=\frac{f(x+\varepsilon)-f(x-\varepsilon)}{2\varepsilon}+\mathcal{O}(\varepsilon^2),$$ and on the stochastic gradient descent. More precisely, we choose $\Delta$ a random vector composed of $-1$ and $1$, each appearing with probability $1/2$, and we approximate our gradient in the direction $\Delta$ as $$\nabla f(\theta)\approx \frac{f(\theta+\varepsilon\Delta)-f(\theta-\varepsilon\Delta)}{2\varepsilon}\Delta$$ since the element-wise inverse of $\Delta$ is itself. Then we update the parameters $\theta$ as usual: $\theta\leftarrow \theta-\eta\nabla f(\theta)$. We choose a batch size of 32, because too little batch might not succeed to capture the global minimum of the landscape. 
We gather in the table in Table \ref{fig:tab_acc_QNN_layers_SPSB} the same results for the two different datasets. \\

\begin{table}[!h]
    \centering
    \begin{tabular}{|c|c|c|c|c|c|c|}
  \hline
    qubits/layers & 4/1 & 4/2 & 4/3  \\
  \hline
    OIF (train/test) & 0.73/0.725 & 0.567/0.62 & 0.641/0.657   \\
    Images (train/set) & 0.552/0.53 & 0.457/0.4315 & 0.572/0.5985  \\
  \hline
  qubits/layers & 7/1 & 7/2 & 7/3 \\
  \hline 
  OIF (train/test) & 0.441/0.635 & 0.503/0.48 & 0.81/0.809 \\
  Images (train/test) & 0.521/0.501 & 0.548/0.569 & 0.535/0.534 \\
  \hline
\end{tabular}\\
    \caption{Accuracy scores of the QNN with SPSB method, for different numbers of qubits and different numbers of layers, on both datasets.}
    \label{fig:tab_acc_QNN_layers_SPSB}
\end{table}

We run the same set of tests. Thus the results for different learning rate values are gathered in Table \ref{fig:tab_acc_QNN_learning_SPSB}, the accuracy scores for the quantum circuit composed of 4 qubits, 10 layers and 10 epochs in Table \ref{fig:tab_acc_QNN_4/10/10_SPSB} and those of the model with 7 qubits, 5 layers and 5 epochs in Table \ref{fig:tab_acc_QNN_7/5/5_SPSB}. All these tables are postponed in the Appendix. For the dataset OIF, we also run another test, with 7 qubits, 5 layers, 5 epochs and $\eta=0.1$, but this time we take 10 000 samples for the train (and we do not change the size of the test). We report the accuracy scores of each epoch in Table \ref{fig:tab_acc_QNN_7/5/5_SPSB_10K}. We present in Table \ref{fig:tab_acc_QNN_4/5/10_SPSB} the accuracy scores for each epoch of our algorithm with 4 qubits, 5 layers, 10 epochs, and $\eta=0.05$, on the two different datasets. 





\begin{table}[!h]
    \centering
    \begin{tabular}{|c|c|c|c|c|c|c|}
  \hline
     & epoch 1 & epoch 2 & epoch 3 & epoch 4 & epoch 5 \\
  \hline
    OIF & 0.485 & 0.744 & 0.803 & 0.814 & 0.816   \\
    Images & 0.562 & 0.627 & 0.683 & 0.709 & 0.719 \\
  \hline
\end{tabular}\\

\begin{tabular}{|c|c|c|c|c|c|c|}
  \hline
     & epoch 6 & epoch 7 & epoch 8 & epoch 9 & epoch 10 & test \\
  \hline
    OIF & 0.816 & 0.816 & 0.816 & 0.817 & 0.817 & 0.8195  \\
    Images & 0.713 & 0.731 & 0.738 & 0.74 & 0.744 & 0.7275 \\
  \hline
\end{tabular}\\
    \caption{Accuracy scores of the QNN with SPSB method, for each epoch, with 4 qubits and 5 layers, on both datasets.}
    \label{fig:tab_acc_QNN_4/5/10_SPSB}
\end{table}


On these tests, we can see that the results are not very concluding when we work with only one epoch, especially on the dataset OIF where we had good accuracy scores for these first tests. On this dataset, we see an increasing of the accuracy score as the learning rate increases, but this is not confirmed when we look at the accuracy scores for the dataset "Images". Again, it is hard to give conclusions for the influence of the learning rate with only three tests, we could test more values of learning rate if we want to observe a real tendency. 

The main difference with the parameter shift rule method is when we look at deeper circuits. Here, we can see a good improvement of the accuracy through the epochs, whereas for the previous method the accuracy was stagnating after two or three epochs. At the end of the processes, we recover similar scores to the scores obtained with the parameter shift rule. More precisely, while the scores for the dataset OIF are slightly smaller with the SPSB method, they are much better on the dataset "Images". We do not see any real difference between the two circuits using 4 qubits, and the circuit with 7 qubits has similar results on the dataset OIF. However, this last circuit with five layers and five epochs does not perform very well on the dataset "Images", it might be due to the fact that the algorithm needs more epochs to learn correctly on this dataset, because we see no stagnation on the last epochs in this case. On the dataset "feature", the algorithms seem to stagnate around the accuracy score of 0.81, this can be seen with the last table where we increased the size of the train dataset to 10 000, and recover the same accuracy as with a smaller train dataset. This is an interesting result in our attempt to learn enough with little data.

The big advantage of this method is that it does not depend on the number of parameters, thus it is largely faster than the parameter shift rule. For example, the SPSB method can compute the tested deep circuits in a couple of hours, against a couple of days for the first method. Thus, a good strategy seems to be computing deep circuits with sufficiently enough epochs with the SPSB method, so that the algorithm can reach good accuracy scores, because the performances will be similar to the ones that we can obtain with the parameter shift rule, and because it will be faster to train the algorithms.  

\subsection{Data reuploading}
In this section, we investigate the potential advantages to the method called data reuploading. It was first introduced in \cite{P_rez_Salinas_2020} in order to bypass the limitations due to the \textit{no cloning} theorem and to hope for a better accuracy without increasing the number of qubits. This method allows to keep track of the input data in the quantum circuit longer than if we just encode it at the beginning of the quantum circuit. Thus, we do not have a circuit composed of an encoding map and several parameterized layers, but directly several layers composed of parameters encoding the input data and parameters to be optimized. More precisely, consider the quantum circuit described in Figure \ref{Circuit quantique}, but this time the layer of the circuit corresponds to the box which was delimiting the encoding map combined to the box which was delimiting the former layer.
We gather in Table \ref{fig:tab_acc_QNN_layers_reupload} the same results for the two different datasets. 

\begin{table}[!h]
    \centering
    \begin{tabular}{|c|c|c|c|c|c|c|}
  \hline
    qubits/layers & 4/1 & 4/2 & 4/3 \\
  \hline
    OIF (train/test) & 0.698/0.667 & 0.722/0.879 & 0.637/0.64   \\
    Images (train/set) & 0.5/0.518 & 0.63/0.52175 & 0.677/0.75325  \\
  \hline
  qubits/layers & 7/1 & 7/2 & 7/3 \\
  \hline
  OIF (train/test) & 0.697/0.784 & 0.805/0.80725 & 0.804/0.86925 \\
  Images (train/test) & 0.513/0.497 & 0.549/0.74775 & 0.602/0.52925 \\
  \hline
\end{tabular}\\
    \caption{Accuracy scores of the QNN with data reuploading method, for different numbers of qubits and different numbers of layers, on both datasets}
    \label{fig:tab_acc_QNN_layers_reupload}
\end{table}

As for the previous methods, the results for different learning rate values are gathered in Table \ref{fig:tab_acc_QNN_learning_reupload}, the accuracy scores for the quantum circuit composed of 4 qubits, 10 layers and 10 epochs in Table \ref{fig:tab_acc_QNN_4/10/10_reupload} and those of the model with 7 qubits, 5 layers and 5 epochs in Table \ref{fig:tab_acc_QNN_7/5/5_reupload}. All these tables are postponed in the Appendix. We present in Table \ref{fig:tab_acc_QNN_4/5/10_reupload} the accuracy scores for each epoch of our algorithm with 4 qubits, 5 layers, 10 epochs, and $\eta=0.05$, on the two different datasets.




\begin{table}[!h]
    \centering
    \begin{tabular}{|c|c|c|c|c|c|c|}
  \hline
     & epoch 1 & epoch 2 & epoch 3 & epoch 4 & epoch 5 \\
  \hline
    OIF & 0.802 & 0.86 & 0.858 & 0.86 & 0.86   \\
    Images & 0.749 & 0.767 & 0.759 & 0.755 & 0.75 \\
  \hline
\end{tabular}\\

\begin{tabular}{|c|c|c|c|c|c|c|}
  \hline
     & epoch 6 & epoch 7 & epoch 8 & epoch 9 & epoch 10 & test \\
  \hline
    OIF & 0.863 & 0.863 & 0.864 & 0.863 & 0.862 & 0.8705  \\
    Images & 0.755 & 0.754 & 0.754 & 0.754 & 0.753 & 0.7095 \\
  \hline
\end{tabular}
    \caption{Accuracy scores of the QNN with data reuploading method, for each epoch, with 4 qubits and 5 layers, on both datasets.}
    \label{fig:tab_acc_QNN_4/5/10_reupload}
\end{table}

With this method, we can observe better results from the initial tests, namely with only one epoch. It is even more noticeable on the dataset "Images". Concerning the influence of the learning rate, the algorithms seem to perform better for bigger learning rates, but we only run a few tests to track this dependence. 
When we look at the deeper circuits, here we observe, as for the parameter shift rule without data reuploading, that the algorithms seem not to learn a lot after two or three epochs. However, the data reuploading method gives better results, which are much more similar to the scores of the classical neural network. Here again, the two algorithms with 4 qubits give similar results, which might indicate that the overparameterisation regime is already reached, and thus it is not necessary to add more layers. We do not observe any real advantage in increasing the number of qubits. 

With data reuploading, the quantum circuits become automatically deeper, because we have to re-encode the input data several time, with multiple quantum gates. However, these gates do not depend on the trained parameters so they are not impacted when computing the parameter shift rule. Thus, the computation times for this method are slightly longer, but still comparable, than the ones of  first method, where we used the parameter shift rule without data reuploading. However the gain in accuracy is non-negligible, and allows to get closer to the results of a classical neural network.

Considering the last two methods, it could be interesting to combine both SPSB and data reuploading in order to obtain good accuracy results with a reduced computation time. This is investigated in the next section.

\subsection{Data reuploading combined to SPSB}
In this last section, we test if the combination of both of theses techniques yields a better result than using only one. 
We still choose a batch size of 32, because too little batch might not succeed to capture the global minimum of the landscape. It would be interesting to observe the same behaviour as when we just set up SPSB method, because it would allow to get similar results to those obtained with just data reuploading method, but with a largely faster model.
We gather in Table \ref{fig:tab_acc_QNN_layers_SPSB_reupload} the same results for the two different datasets. 

\begin{table}[!h]
    \centering
    \begin{tabular}{|c|c|c|c|c|c|c|}
  \hline
    qubits/layers & 4/1 & 4/2 & 4/3  \\
  \hline
    OIF (train/test) & 0.73/0.725 & 0.561/0.61075 & 0.468/0.479  \\
    Images (train/set) & 0.552/0.53 & 0.46/0.46 & 0.49/0.5005  \\
  \hline
  qubits/layers & 7/1 & 7/2 & 7/3 \\
  \hline
  OIF (train/test) & 0.441/0.635 & 0.709/0.77075 & 0.432/0.514 \\
  Images (train/test) & 0.521/0.501 & 0.545/0.54625 & 0.512/0.54125 \\
  \hline
\end{tabular}\\
    \caption{Accuracy scores of the QNN with data reuploading and SPSB methods, for different numbers of qubits and different numbers of layers, on both datasets.}
    \label{fig:tab_acc_QNN_layers_SPSB_reupload}
\end{table}

As in the previous sections, we gather in Table \ref{fig:tab_acc_QNN_learning_SPSB_reupload} the results for different learning rate values. Tables \ref{fig:tab_acc_QNN_4/10/10_SPSB_reupload} and \ref{fig:tab_acc_QNN_7/5/5_SPSB_reupload} respectively present the accuracy scores for the quantum circuit composed of 4 qubits, 10 layers and 10 epochs, and for the model with 7 qubits, 5 layers and 5 epochs.
All these tables are postponed in the Appendix. We present in Table \ref{fig:tab_acc_QNN_4/5/10_SPSB_reupload} the accuracy scores for each epoch of our algorithm with 4 qubits, 5 layers, 10 epochs, and $\eta=0.05$, on the two different datasets.


\begin{table}[!h]
    \centering
    \begin{tabular}{|c|c|c|c|c|c|c|}
  \hline
     & epoch 1 & epoch 2 & epoch 3 & epoch 4 & epoch 5 \\
  \hline
    OIF & 0.467 & 0.466 & 0.555 & 0.675 & 0.834   \\
    Images & 0.575 & 0.645 & 0.686 & 0.725 & 0.735 \\
  \hline
\end{tabular}\\

\begin{tabular}{|c|c|c|c|c|c|c|}
  \hline
     & epoch 6 & epoch 7 & epoch 8 & epoch 9 & epoch 10 & test \\
  \hline
    OIF & 0.843 & 0.848 & 0.857 & 0.856 & 0.865 & 0.864  \\
    Images & 0.736 & 0.742 & 0.746 & 0.753 & 0.746 & 0.75425 \\
  \hline
\end{tabular}\\
    \caption{Accuracy scores of the QNN with data reuploading and SPSB methods, for each epoch, with 4 qubits and 5 layers, on both datasets.}
    \label{fig:tab_acc_QNN_4/5/10_SPSB_reupload}
\end{table}


In the tests of this last method, we can observe the same behaviour as for the method where just SPSB was used, which is good news. Indeed, for the circuits with only one epoch, the models exhibit no noticeable improvement. So in this case the data reuploading method is not really efficient, or at least it doesn't compensate the lack of accuracy due to SPSB. We again observe an increasing of the accuracy as the learning rate increases, but since we run the tests for only three learning rate values again it is not very relevant.

On deeper circuits, we can see that the models learn through the epochs, and it seems to need a little more epochs to stabilize than when we used only SPSB. This is shown especially for the model with 7 qubits, 5 layers and 5 epochs, Table \ref{fig:tab_acc_QNN_7/5/5_SPSB_reupload}, where we do not get scores as good as for the others models. In this case the learning seems to be incomplete, in the sense that adding more epochs would probably yield better results (we can see that the accuracy is still increasing, and is not stabilized yet). However, the two others models, namely the circuits with 4 qubits, we obtain accuracy scores which are similar to those obtained with the method using data reuploading and parameter shift rule. Considering the computation times of the two methods, it is thus a great advantage to be able to recover such scores with a largely faster method. This time the circuit with 10 layers seems to perform slightly better than the one with 5 layers. 

Note that for the methods using the SPSB, it is not very constraining to add epochs, because the gain in time computation compared to parameter shift rule is very important. Thus, if one needs to run a model with for example 10 layers rather than 5, the use of SPSB reasonably allows it. 
The combination of this method with data reuploading is a good alternative for deep circuits, because it dramatically reduces the time computation and gives better results than a method without reuploading.

\section{Conclusion and future work}

In this paper, we observed the performances of different models for two types of preprocessings. For each model, the preprocessing consisting in extracting the most impactful features gave much better results than the Grayscale method, surely because the algorithms are more adapted to statistical or semantic features than to images. Using a PCA with so few components on this particular type of images also makes us lose a lot of information.

We compared our quantum models with classical equivalent ones. The first quantum model, i.e the QSVM, seems to give satisfying results on real quantum hardware, and gives for some parameters better results than several classical SVM when computed on quantum simulator. On the other hand, the QNN never gives better results than the classical NN, even if its performances get close to the performances of the classical model when we optimize it. Indeed, we used two methods in order to get better results with our QNN. While we have no optimal architecture lead yet, we implemented data reuploading on our model, which re-encode the input data several times in the same quantum circuit, to keep track on the information contained in it. This method allowed to increase the accuracy scores of the QNN, becoming quite comparable to those of a classical neural network. However this method increases the depth of the quantum circuits, and this inconvenient can raise issues in term of computation time when one uses the parameter shift rule method for the gradient descent. Thus we also investigated the SPSB method, which allows a largely faster gradient descent. We observed that for a sufficiently large number of epochs, namely when the model has "enough time to learn correctly", the algorithm reaches similar results as the algorithm with parameter shift rule. It is thus a good alternative, because the number of computations of the quantum circuit in this method does not depend on the number of trainable parameters in the model. We conclude that the combination of these two methods is a fine compromise, allowing to get results that are comparable to a classical neural network, with a computation time which is drastically smaller than for the usual parameter shift rule. 

In order to get even better results, several directions can be considered. First, we mentioned in our observations the notion of overparameterization. This is studied in \cite{Larocca_2021}, where the authors describe this phenomenon as the regime where the quantum model has sufficiently many parameters so that the Fisher Information matrix reaches its maximal rank, and thus has maximal capacity. Such a phenomenon is linked to the dynamical Lie algebra obtained from the set of generators of the QNN. It could be useful to identify the dynamical Lie algebra in the case of our model, because it will help us to reach the overparameterization regime, and thus we will know for sure that it is no more necessary to add layers of parameters to our quantum circuit. 
Another lead is to work on the input data, in order to identify patterns such as symmetries in it. For instance it could be the fact that swapping two features  does not change the learning. This could allow to lead the construction of an appropriate quantum circuit, called equivariant circuit, as presented in  \cite{Meyer2022} \cite{Ragone_2022}. This type of quantum model can respect the symmetries of the data and gives good results with less parameters, hence less computations if one thinks for example to methods such as parameter shift rule. 
Note also that in this work, we do not discuss about the number of shots used for the estimates of the gradients, which might be optimized in order to reduce the computation time without loss of performance. In \cite{moussa2022resource}, the authors explain the importance of this discussion, and propose an unbiased estimator for the gradients of a general loss function, allowing a shot frugal gradient descent. 
Finally, we could also work on a different model, such as a Quantum Convolutional Neural Network (QCNN). It is proved that QCNNs do not experiment Barren plateaus \cite{PhysRevX.11.041011}, and such models can be used for classification problem \cite{Bokhan_2022}\cite{Hur_2022}, and potentially perform better than the quantum models studied in our work, especially on the dataset composed of images.

\bibliographystyle{plain}
\bibliography{biblio}

\newpage
\section*{Appendix}

In this appendix we gather some results of tests that we didn't present directly in the article in order to lighten the lecture of the other sections. This time we choose to present theses results regrouped according to the type of test rather than according to the type of method used. 

Concerning our model of QNN, we tried to illustrate the influence of the learning rate on the accuracy of the model. However, we just tested our algorithms for three learning rate values, namely $\eta=0.01$, $\eta=0.1$ and $\eta=0.5$. Thus, it is hard to really extract a global behaviour from our results. Nevertheless, we present several tables of accuracy scores, namely Table \ref{fig:tab_acc_QNN_learning_img200} to Table \ref{fig:tab_acc_QNN_learning_SPSB_reupload}. All the tests were made with one epoch and two layers. Table \ref{fig:tab_acc_QNN_learning_img200} corresponds to the dataset "Images" with 1000 training samples and 200 test samples. Table \ref{fig:tab_acc_QNN_learning_img} corresponds to the same dataset, but this time with 4000 test samples. Then we keep 1000 training samples and 4000 test samples. Table \ref{fig:tab_acc_QNN_learning_features} corresponds to the dataset OIF. Finally, we add three other tables corresponding to both datasets, but respectively with SPSB (Table \ref{fig:tab_acc_QNN_learning_SPSB}), data reuploading (Table \ref{fig:tab_acc_QNN_learning_reupload}), then SPSB and data reuploading (Table \ref{fig:tab_acc_QNN_learning_SPSB_reupload}) applied on the models.  Tables \ref{fig:tab_acc_NN_learning_img200}, \ref{fig:tab_acc_NN_learning_img} and \ref{fig:tab_acc_NN_learning_features} are respective equivalents on a classical neural network of Tables \ref{fig:tab_acc_QNN_learning_img200}, \ref{fig:tab_acc_QNN_learning_img} and \ref{fig:tab_acc_QNN_learning_features}.  

Once we have run all of these tests, we implemented deeper circuits. We start by presenting a QNN with 4 qubits, 10 layers and 10 epochs, with a learning rate of $0.05$. Tables corresponding to Tables \ref{fig:tab_acc_QNN_4/10/10_img200} to \ref{fig:tab_acc_QNN_4/10/10_SPSB_reupload} represent the accuracy scores of this model in the same order of frameworks as for the previous tables. Tables \ref{fig:tab_acc_NN_4/10/10_img200}, \ref{fig:tab_acc_NN_4/10/10_img} and \ref{fig:tab_acc_NN_4/10/10_features} are respective equivalents on a classical neural network of Tables \ref{fig:tab_acc_QNN_4/10/10_img200}, \ref{fig:tab_acc_QNN_4/10/10_img} and \ref{fig:tab_acc_QNN_4/10/10_features}.

Finally, we present a last model, with a QNN composed of 7 qubits, 5 layers and 5 epochs, with a learning rate of $0.1$. Here again, the tables corresponding to Tables \ref{fig:tab_acc_QNN_7/5/5_img200} to \ref{fig:tab_acc_QNN_7/5/5_SPSB_reupload} represent the accuracy scores of this model in the same order of frameworks as for the previous tables. Tables \ref{fig:tab_acc_NN_7/5/5_img200}, \ref{fig:tab_acc_NN_7/5/5_img} and \ref{fig:tab_acc_NN_7/5/5_features} are respective equivalents on a classical neural network of Tables \ref{fig:tab_acc_QNN_7/5/5_img200}, \ref{fig:tab_acc_QNN_7/5/5_img} and \ref{fig:tab_acc_QNN_7/5/5_features}, and when it comes to the QNN on the dataset OIF with the SPSB method, we add a series of tests for 10 000 training samples and 4000 test samples (Table \ref{fig:tab_acc_QNN_7/5/5_SPSB_10K}).

\begin{table}[!h]
    \centering
    \begin{tabular}{|c|c|c|c|}
  \hline
    qubits/layers & $\eta=0.01$ & $\eta=0.1$ & $\eta=0.5$ \\
  \hline
    4/2 & 0.5/0.525 & 0.609/0.605 & 0.552/0.502 \\
    7/2 & 0.548/0.475 & 0.513/0.445 & 0.491/0.445 \\
  \hline
\end{tabular}\\
    \caption{Accuracy scores of the QNN for different numbers of qubits and different values of learning rates, on the dataset "Binaires.image" for 200 test samples. }
    \label{fig:tab_acc_QNN_learning_img200}
\end{table}

\begin{table}[!h]
    \centering
    \begin{tabular}{|c|c|c|c|}
  \hline
    PCA/layers & $\eta=0.01$ & $\eta=0.1$ & $\eta=0.5$ \\
  \hline
    4/2 & 0.5040/0.5850 & 0.489/0.410 & 0.5050/0.420 \\
    7/2 & 0.5040/0.5950 & 0.489/0.4350 & 0.5050/0.580 \\
  \hline
\end{tabular}\\
    \caption{Accuracy scores of the classical NN for different PCA values and different learning rate values, on the dataset "Images" for 200 test samples.}
    \label{fig:tab_acc_NN_learning_img200}
\end{table}

\begin{table}[!h]
    \centering
    \begin{tabular}{|c|c|c|c|}
  \hline
    qubits/layers & $\eta=0.01$ & $\eta=0.1$ & $\eta=0.5$ \\
  \hline
    4/2 & 0.49/0.495 & 0.515/0.508 & 0.517/0.509 \\
    7/2 & 0.543/0.565 & 0.537/0.5077 & 0.527/0.509 \\
  \hline
\end{tabular}\\
    \caption{Accuracy scores of the QNN, for different numbers of qubits and different learning rate values, on the dataset "Images".}
    \label{fig:tab_acc_QNN_learning_img}
\end{table}

\begin{table}[!h]
    \centering
    \begin{tabular}{|c|c|c|c|}
  \hline
    PCA/layers & $\eta=0.01$ & $\eta=0.1$ & $\eta=0.5$ \\
  \hline
    4/2 & 0.748/0.639 & 0.668/0.630 & 0.630/0.6695 \\
    7/2 & 0.740/0.7775 & 0.663/0.5573 & 0.634/0.5228 \\
  \hline
\end{tabular}
    \caption{Accuracy scores of the classical NN, for different PCA values and different learning rate values, on the dataset "Images".}
    \label{fig:tab_acc_NN_learning_img}
\end{table}

\begin{table}[!h]
    \centering
    \begin{tabular}{|c|c|c|c|}
  \hline
    qubits/layers & $\eta=0.01$ & $\eta=0.1$ & $\eta=0.5$ \\
  \hline
    4/2 & 0.641/0.6695 & 0.677/0.667 & 0.687/0.839 \\
    7/2 & 0.503/0.482 & 0.487/0.504 & 0.637/0.80025 \\
  \hline
\end{tabular}\\
    \caption{Accuracy scores of the QNN, for different numbers of qubits and different learning rate values, on the dataset OIF.}
    \label{fig:tab_acc_QNN_learning_features}
\end{table}

\begin{table}[!h]
    \centering
    \begin{tabular}{|c|c|c|c|}
  \hline
    features/layers & $\eta=0.01$ & $\eta=0.1$ & $\eta=0.5$ \\
  \hline
    4/2 & 0.85/0.711 & 0.8370/0.5910 & 0.8120/0.4695 \\
    7/2 & 0.8540/0.8953 & 0.829/0.5312 & 0.8060/0.4715 \\
  \hline
\end{tabular}\\
    \caption{Accuracy scores of the classical NN, for different numbers of features and different learning rate values, on the dataset OIF.}
    \label{fig:tab_acc_NN_learning_features}
\end{table}

\begin{table}[!h]
    \centering
    \begin{tabular}{|c|c|c|c|}
  \hline
    qubits/layers & $\eta=0.01$ & $\eta=0.1$ & $\eta=0.5$ \\
  \hline
    4/2 (OIF) & 0.482/0.485 & 0.567/0.62 & 0.672/0.80 \\
    7/2 (OIF) & 0.503/0.4805 & 0.503/0.48 & 0.506/0.4785 \\
  \hline
    4/2 (Images) & 0.462/0.439 & 0.457/0.4315 & 0.484/0.531 \\
    7/2 (Images) & 0.548/0.5705 & 0.548/0.569 & 0.544/0.571 \\
  \hline
\end{tabular}\\
    \caption{Accuracy scores of the QNN with SPSB method, for different number of qubits and different learning rate values, on both datasets.}
    \label{fig:tab_acc_QNN_learning_SPSB}
\end{table}

\begin{table}[!h]
    \centering
    \begin{tabular}{|c|c|c|c|}
  \hline
    qubits/layers & $\eta=0.01$ & $\eta=0.1$ & $\eta=0.5$ \\
  \hline
    4/2 (OIF) & 0.641/0.6685 & 0.722/0.879 & 0.748/0.8715 \\
    7/2 (OIF) & 0.767/0.804 & 0.805/0.80725 & 0.803/0.80775 \\
  \hline
    4/2 (Images) & 0.509/0.51975 & 0.63/0.52175 & 0.577/0.48925 \\
    7/2 (Images) & 0.534/0.53675 & 0.549/0.74775 & 0.516/0.5085 \\
  \hline
\end{tabular}\\
    \caption{Accuracy scores of the QNN with data reuploading method, for different numbers of qubits and different learning rate values, on both datasets.}
    \label{fig:tab_acc_QNN_learning_reupload}
\end{table}

\begin{table}[!h]
    \centering
    \begin{tabular}{|c|c|c|c|}
  \hline
    qubits/layers & $\eta=0.01$ & $\eta=0.1$ & $\eta=0.5$ \\
  \hline
    4/2(OIF) & 0.478/0.48275 & 0.561/0.61075 & 0.655/0.7935 \\
    7/2(OIF) & 0.643/0.645 & 0.709/0.77075 & 0.77/0.80775 \\
  \hline
    4/2(Images) & 0.483/0.4605 & 0.46/0.46 & 0.537/0.752 \\
    7/2(Images) & 0.543/0.54 & 0.545/0.54625 & 0.561/0.578 \\
  \hline
\end{tabular}\\
    \caption{Accuracy scores of the QNN with data reuploading and SPSB methods, for different numbers of qubits and different learning rate values, on both datasets.}
    \label{fig:tab_acc_QNN_learning_SPSB_reupload}
\end{table}

\begin{table}[!h]
    \centering
    \begin{tabular}{|c|c|c|c|c|c|c|}
  \hline
     & epoch 1 & epoch 2 & epoch 3 & epoch 4 & epoch 5 \\
  \hline
    4 qubits/10layers & 0.633 & 0.667 & 0.666 & 0.675 & 0.681   \\
  \hline
\end{tabular}\\
\begin{tabular}{|c|c|c|c|c|c|c|}
  \hline
     & epoch 6 & epoch 7 & epoch 8 & epoch 9 & epoch 10 & test \\
  \hline
    4 qubits/10layers & 0.682 & 0.682 & 0.678 & 0.679 & 0.674 & 0.755  \\
  \hline
\end{tabular}\\
    \caption{Accuracy scores of the QNN, for each epoch, with 4 qubits and 10 layers, on the dataset "Images" for 200 test samples. }
    \label{fig:tab_acc_QNN_4/10/10_img200}
\end{table}


\begin{table}[!h]
    \centering
    \begin{tabular}{|c|c|c|c|c|c|c|}
  \hline
     & epoch 1 & epoch 2 & epoch 3 & epoch 4 & epoch 5 \\
  \hline
    PCA 4/10layers & 0.5020 & 0.5170 & 0.476 & 0.4930 & 0.5170   \\
  \hline
\end{tabular}\\

\begin{tabular}{|c|c|c|c|c|c|c|}
  \hline
     & epoch 6 & epoch 7 & epoch 8 & epoch 9 & epoch 10 & test \\
  \hline
    PCA 4/10layers & 0.5200 & 0.4870 & 0.5050 & 0.5010 & 0.4880 & 0.5400  \\
  \hline
\end{tabular}\\
    \caption{Accuracy scores of the classical NN, for each epoch, with PCA 4 and 10 layers, on the dataset "Images" for 200 test samples.}
    \label{fig:tab_acc_NN_4/10/10_img200}
\end{table}

\begin{table}[!h]
    \centering
    \begin{tabular}{|c|c|c|c|c|c|c|}
  \hline
     & epoch 1 & epoch 2 & epoch 3 & epoch 4 & epoch 5 \\
  \hline
    4 qubits/10layers & 0.525 & 0.56 & 0.57 & 0.57 & 0.56   \\
  \hline
\end{tabular}\\

\begin{tabular}{|c|c|c|c|c|c|c|}
  \hline
     & epoch 6 & epoch 7 & epoch 8 & epoch 9 & epoch 10 & test \\
  \hline
    4 qubits/10layers & 0.57 & 0.57 & 0.57 & 0.57 & 0.57 & 0.53  \\
  \hline
\end{tabular}\\
    \caption{Accuracy scores of the QNN, for each epoch, with 4 qubits and 10 layers, on the dataset "Images"}
    \label{fig:tab_acc_QNN_4/10/10_img}
\end{table}


\begin{table}[!h]
    \centering
    \begin{tabular}{|c|c|c|c|c|c|c|}
  \hline
     & epoch 1 & epoch 2 & epoch 3 & epoch 4 & epoch 5 \\
  \hline
    PCA 4/10layers & 0.64 & 0.732 & 0.73 & 0.74 & 0.76   \\
  \hline
\end{tabular}\\

\begin{tabular}{|c|c|c|c|c|c|c|}
  \hline
     & epoch 6 & epoch 7 & epoch 8 & epoch 9 & epoch 10 & test \\
  \hline
    PCA 4/10layers & 0.74 & 0.75 & 0.76 & 0.75 & 0.74 & 0.49  \\
  \hline
\end{tabular}\\
    \caption{Accuracy scores of the classical NN, for each epoch, with PCA 4 and 10 layers, on the dataset "Images".}
    \label{fig:tab_acc_NN_4/10/10_img}
\end{table}

\begin{table}[!h]
    \centering
    \begin{tabular}{|c|c|c|c|c|c|c|}
  \hline
     & epoch 1 & epoch 2 & epoch 3 & epoch 4 & epoch 5 \\
  \hline
    4 qubits/10layers & 0.812 & 0.812 & 0.812 & 0.812 & 0.812   \\
  \hline
\end{tabular}\\

\begin{tabular}{|c|c|c|c|c|c|c|}
  \hline
     & epoch 6 & epoch 7 & epoch 8 & epoch 9 & epoch 10 & test \\
  \hline
    4 qubits/10layers & 0.812 & 0.812 & 0.812 & 0.812 & 0.812 & 0.8135  \\
  \hline
\end{tabular}\\
    \caption{Accuracy scores of the QNN, for each epoch, with 4 qubits and 10 layers, on the dataset OIF.}
    \label{fig:tab_acc_QNN_4/10/10_features}
\end{table}


\begin{table}[!h]
    \centering
    \begin{tabular}{|c|c|c|c|c|c|c|}
  \hline
     & epoch 1 & epoch 2 & epoch 3 & epoch 4 & epoch 5 \\
  \hline
    4 features/10layers & 0.8560 & 0.8720 & 0.8700 & 0.8810 & 0.8890   \\
  \hline
\end{tabular}\\

\begin{tabular}{|c|c|c|c|c|c|c|}
  \hline
     & epoch 6 & epoch 7 & epoch 8 & epoch 9 & epoch 10 & test \\
  \hline
    4 features/10layers & 0.8640 & 0.8910 & 0.8760 & 0.8780 & 0.8810 & 0.8900  \\
  \hline
\end{tabular}\\
    \caption{Accuracy scores of the classical NN, for each epoch, with 4 features and 10 layers, on the dataset OIF.}
    \label{fig:tab_acc_NN_4/10/10_features}
\end{table}

\begin{table}[!h]
    \centering
    \begin{tabular}{|c|c|c|c|c|c|c|}
  \hline
     & epoch 1 & epoch 2 & epoch 3 & epoch 4 & epoch 5 \\
  \hline
    OIF & 0.555 & 0.808 & 0.808 & 0.808 & 0.808   \\
    Images & 0.516 & 0.521 & 0.519 & 0.515 & 0.56 \\
  \hline
\end{tabular}\\

\begin{tabular}{|c|c|c|c|c|c|c|}
  \hline
     & epoch 6 & epoch 7 & epoch 8 & epoch 9 & epoch 10 & test \\
  \hline
    OIF & 0.809 & 0.809 & 0.811 & 0.811 & 0.815 & 0.8167  \\
    Images & 0.732 & 0.745 & 0.736 & 0.743 & 0.728 & 0.67925 \\
  \hline
\end{tabular}\\
    \caption{Accuracy scores of the QNN with SPSB method, for each epoch, with 4 qubits and 10 layers, on both datasets.}
    \label{fig:tab_acc_QNN_4/10/10_SPSB}
\end{table}

\begin{table}[!h]
    \centering
    \begin{tabular}{|c|c|c|c|c|c|c|}
  \hline
     & epoch 1 & epoch 2 & epoch 3 & epoch 4 & epoch 5 \\
  \hline
    OIF & 0.874 & 0.885 & 0.882 & 0.88 & 0.883   \\
    Images & 0.76 & 0.756 & 0.76 & 0.754 & 0.753 \\
  \hline
\end{tabular}\\

\begin{tabular}{|c|c|c|c|c|c|c|}
  \hline
     & epoch 6 & epoch 7 & epoch 8 & epoch 9 & epoch 10 & test \\
  \hline
    OIF & 0.879 & 0.879 & 0.883 & 0.877 & 0.873 & 0.88175  \\
    Images & 0.753 & 0.754 & 0.757 & 0.76 & 0.759 & 0.77 \\
  \hline
\end{tabular}\\
    \caption{Accuracy scores of the QNN with data reuploading method, for each epoch, with 4 qubits and 10 layers, on both datasets.}
    \label{fig:tab_acc_QNN_4/10/10_reupload}
\end{table}

\begin{table}[!h]
    \centering
    \begin{tabular}{|c|c|c|c|c|c|c|}
  \hline
     & epoch 1 & epoch 2 & epoch 3 & epoch 4 & epoch 5 \\
  \hline
    OIF & 0.366 & 0.699 & 0.827 & 0.841 & 0.841   \\
    Images & 0.582 & 0.691 & 0.728 & 0.739 & 0.763 \\
  \hline
\end{tabular}\\

\begin{tabular}{|c|c|c|c|c|c|c|}
  \hline
     & epoch 6 & epoch 7 & epoch 8 & epoch 9 & epoch 10 & test \\
  \hline
    OIF & 0.84 & 0.856 & 0.856 & 0.874 & 0.874 & 0.876  \\
    Images & 0.752 & 0.769 & 0.768 & 0.765 & 0.764 & 0.76425 \\
  \hline
\end{tabular}\\
    \caption{Accuracy scores of the QNN with data reuploading and SPSB methods, for each epoch, with 4 qubits and 10 layers, on both datasets.}
    \label{fig:tab_acc_QNN_4/10/10_SPSB_reupload}
\end{table}

\begin{table}[!h]
    \centering
    \begin{tabular}{|c|c|c|c|c|c|c|}
  \hline
     & epoch 1 & epoch 2 & epoch 3 & epoch 4 & epoch 5 & test \\
  \hline
    7 qubits/5layers & 0.61 & 0.54 & 0.58 & 0.615 & 0.58 & 0.55  \\
  \hline
\end{tabular}\\
    \caption{Accuracy scores of the QNN, for each epoch, with 7 qubits and 5 layers, on the dataset "Images" for 200 test samples.}
    \label{fig:tab_acc_QNN_7/5/5_img200}
\end{table}


\begin{table}[!h]
    \centering
    \begin{tabular}{|c|c|c|c|c|c|c|}
  \hline
     & epoch 1 & epoch 2 & epoch 3 & epoch 4 & epoch 5 & test \\
  \hline
    PCA 7/5layers & 0.4860 & 0.5190 & 0.4790 & 0.5120 & 0.5190 & 0.6000  \\
  \hline
\end{tabular}\\
    \caption{Accuracy scores of the classical NN, for each epoch, with PCA 7 and 5 layers, on the dataset "Images" for 200 test samples.}
    \label{fig:tab_acc_NN_7/5/5_img200}
\end{table}

\begin{table}[!h]
    \centering
    \begin{tabular}{|c|c|c|c|c|c|c|}
  \hline
     & epoch 1 & epoch 2 & epoch 3 & epoch 4 & epoch 5 & test \\
  \hline
    7 qubits/5layers & 0.517 & 0.526 & 0.521 & 0.538 & 0.544 & 0.5425  \\
  \hline
\end{tabular}\\
    \caption{Accuracy scores of the QNN, for each epoch, with 7 qubits and 5 layers, on the dataset "Images".}
    \label{fig:tab_acc_QNN_7/5/5_img}
\end{table}


\begin{table}[!h]
    \centering
    \begin{tabular}{|c|c|c|c|c|c|c|}
  \hline
     & epoch 1 & epoch 2 & epoch 3 & epoch 4 & epoch 5 & test \\
  \hline
    PCA 7/5layers & 0.712 & 0.747 & 0.764 & 0.788 & 0.7690 & 0.7845  \\
  \hline
\end{tabular}\\
    \caption{Accuracy scores of the classical NN, for each epoch, with PCA 7 and 5 layers, on the dataset "Images".}
    \label{fig:tab_acc_NN_7/5/5_img}
\end{table}

\begin{table}[!h]
    \centering
    \begin{tabular}{|c|c|c|c|c|c|c|}
  \hline
     & epoch 1 & epoch 2 & epoch 3 & epoch 4 & epoch 5 & test \\
  \hline
    7 qubits/5layers & 0.811 & 0.811 & 0.812 & 0.816 & 0.818 & 0.815  \\
  \hline
\end{tabular}\\
    \caption{Accuracy scores of the QNN, for each epoch, with 7 qubits and 5 layers, on the dataset OIF.}
    \label{fig:tab_acc_QNN_7/5/5_features}
\end{table}


\begin{table}[!h]
    \centering
    \begin{tabular}{|c|c|c|c|c|c|c|}
  \hline
     & epoch 1 & epoch 2 & epoch 3 & epoch 4 & epoch 5 & test \\
  \hline
    7 features/5layers & 0.832 & 0.87 & 0.885 & 0.874 & 0.87 & 0.889  \\
  \hline
\end{tabular}\\
    \caption{Accuracy scores of the classical NN, for each epoch, with 7 features and 5 layers, on the dataset OIF.}
    \label{fig:tab_acc_NN_7/5/5_features}
\end{table}

\begin{table}[!h]
    \centering
    \begin{tabular}{|c|c|c|c|c|c|c|}
  \hline
     & epoch 1 & epoch 2 & epoch 3 & epoch 4 & epoch 5 & test \\
  \hline
    OIF & 0.462 & 0.729 & 0.807 & 0.809 & 0.81 & 0.8135  \\
    Images & 0.479 & 0.476 & 0.477 & 0.504 & 0.521 & 0.559 \\
  \hline
\end{tabular}\\
    \caption{Accuracy scores of the QNN with SPSB method, for each epoch, with 7 qubits and 5 layers, on both datasets.}
    \label{fig:tab_acc_QNN_7/5/5_SPSB}
\end{table}

\begin{table}[!h]
    \centering
    \begin{tabular}{|c|c|c|c|c|c|c|}
  \hline
     & epoch 1 & epoch 2 & epoch 3 & epoch 4 & epoch 5 & test \\
  \hline
    7 qubits/5layers & 0.7429 & 0.8136 & 0.8153 & 0.8243 & 0.8251 & 0.82025  \\
  \hline
\end{tabular}\\  
    \caption{Accuracy scores of the QNN with SPSB method and 10 000 training samples, for each epoch, with 7 qubits and 5 layers, on the dataset OIF.}
    \label{fig:tab_acc_QNN_7/5/5_SPSB_10K}
\end{table}

\begin{table}[!h]
    \centering
    \begin{tabular}{|c|c|c|c|c|c|c|}
  \hline
     & epoch 1 & epoch 2 & epoch 3 & epoch 4 & epoch 5 & test \\
  \hline
    OIF & 0.782 & 0.862 & 0.869 & 0.873 & 0.873 & 0.881  \\
    Images & 0.62 & 0.656 & 0.73 & 0.737 & 0.736 & 0.71625 \\
  \hline
\end{tabular}\\
    \caption{Accuracy scores of the QNN with data reuploading method, for each epoch, with 7 qubits and 5 layers, on both datasets.}
    \label{fig:tab_acc_QNN_7/5/5_reupload}
\end{table}

\begin{table}[!h]
    \centering
    \begin{tabular}{|c|c|c|c|c|c|c|}
  \hline
     & epoch 1 & epoch 2 & epoch 3 & epoch 4 & epoch 5 & test \\
  \hline
    OIF & 0.619 & 0.703 & 0.743 & 0.762 & 0.777 & 0.7825  \\
    Images & 0.33 & 0.356 & 0.435 & 0.473 & 0.486 & 0.515 \\
  \hline
\end{tabular}\\
    \caption{Accuracy scores of the QNN with data reuploading and SPSB methods, for each epoch, with 7 qubits and 5 layers, on both datasets.}
    \label{fig:tab_acc_QNN_7/5/5_SPSB_reupload}
\end{table}

\end{document}